\documentclass[12pt]{article}
\usepackage{epsfig}

\catcode`\@=11

\newcount\hour
\newcount\minute
\newtoks\amorpm \hour=\time\divide\hour by 60\minute
=\time{\multiply\hour by 60 \global\advance\minute by-\hour}
\edef\standardtime{{\ifnum\hour<12 \global\amorpm={am}%
        \else\global\amorpm={pm}\advance\hour by-12 \fi
        \ifnum\hour=0 \hour=12 \fi
        \number\hour:\ifnum\minute<10
        0\fi\number\minute\the\amorpm}}
\edef\militarytime{\number\hour:\ifnum\minute<10
0\fi\number\minute}

\def\draftlabel#1{{\@bsphack\if@filesw {\let\thepage\relax
   \xdef\@gtempa{\write\@auxout{\string
      \newlabel{#1}{{\@currentlabel}{\thepage}}}}}\@gtempa
   \if@nobreak \ifvmode\nobreak\fi\fi\fi\@esphack}
        \gdef\@eqnlabel{#1}}
\def\@eqnlabel{}
\def\@vacuum{}
\def\marginnote#1{}
\def\draftmarginnote#1{\marginpar{\raggedright\scriptsize\tt#1}}
\def\draft{
        \pagestyle{plain}
        \overfullrule=2pt
        \oddsidemargin -.1truein
        \def\@oddhead{\sl \phantom{\today\quad\militarytime} \hfil
        \smash{\Large\sl DRAFT} \hfil \today\quad\militarytime}
        \let\@evenhead\@oddhead
        \let\label=\draftlabel
        \let\marginnote=\draftmarginnote
        \def\ps@empty{\let\@mkboth\@gobbletwo
        \def\@oddfoot{\hfil \smash{\Large\sl DRAFT} \hfil}
        \let\@evenfoot\@oddhead}
        \def\@eqnnum{(\theequation)\rlap{\kern\marginparsep\tt\@eqnlabel}%
        \global\let\@eqnlabel\@vacuum}  }

\renewcommand{\theequation}{\thesection.\arabic{equation}}
\renewcommand{\thefootnote}{\fnsymbol{footnote}}
\newcommand{\newsection}{    
\setcounter{equation}{0}\section}
\def\appendix#1{\addtocounter{section}{1}\setcounter{equation}{0}
\renewcommand{\thesection}{\Alph{section}}
\section*{Appendix \thesection\protect\indent \parbox[t]{11.15cm}{#1}}
\addcontentsline{toc}{section}{Appendix \thesection\ \ \ #1}}

\def \lc {{light-cone}}
\def \ci{\cite}
\def\f{{\rm f}}

\def \we {\wedge}
\def\nn{\nonumber}
\def \foot{\footnote}
\def \bi{\bibitem}
\def \la {\label}
\def \ov {\over}

\def \x {{\rm x}}
\def \tv {{\rm v}}
\def \four {{1 \ov 4}}
\def \b {\beta}

\def \t {\theta}
\def \s{\sigma}
\def \d {\partial}

\def \del{\partial}

\def \four{{\textstyle {1\ov 4}}}

\def\det{\hbox{det}}

\def\be{\begin{equation}}
\def\ee{\end{equation}}

\def \ci {\cite}

\begin{document}


\date{January 2003}

\begin{titlepage}

\begin{center}
{~}
\vspace{2.5cm}
\vskip 2.0cm
{\Large \bf Anomaly, Fluxes and (2,0) Heterotic-String
 Compactifications \\[.2cm]

%
}

\vspace{1.0cm}
 {\large J. Gillard, G. Papadopoulos and D. Tsimpis}
 \\
 \vspace{1.0cm}
 Department of Mathematics \\
 King's College London\\
 Strand\\
 London WC2R 2LS\\
\end{center}

\vskip 0.4cm

\begin{center}
{\sl To the memory of Sonia
Stanciu   
}
\end{center}

\vskip 2.0 cm

\begin{abstract}

\noindent
We compute
the corrections to
heterotic-string backgrounds with (2,0) world-sheet
supersymmetry, up to two loops in
sigma-model perturbation theory. We investigate
the conditions for these backgrounds to preserve
spacetime supersymmetry and we find that a sufficient requirement
for consistency is
 the applicability
of the $\partial\bar\partial$-lemma. In particular, we
investigate the $\alpha'$ corrections to (2,0) heterotic-string
compactifications and we find that the Calabi-Yau geometry of the
internal space is deformed to a Hermitian one. We show that at
first order in $\alpha'$, the heterotic anomaly-cancellation mechanism
does not induce any lifting of  moduli. We explicitly compute the
corrections to the conifold and 
to the $U(n)$-invariant Calabi-Yau metric at
first order in $\alpha'$. We also find a generalization  of the 
gauge-field equations, compatible with the Donaldson equations on
 conformally-balanced Hermitian manifolds.
\end{abstract}

\end{titlepage}

\newpage
\setcounter{page}{1}
\renewcommand{\thefootnote}{\arabic{footnote}}
\setcounter{footnote}{0}

 \def \lc {light-cone\ }

\def \ae {{\rm E}}
\def \Epsilon {{\cal E}}
\hoffset=35pt
\voffset=-1.5cm
\textwidth=15.8cm
\textheight=23cm

\hoffset=-25pt
\voffset=-2.5cm
\catcode`\@=11

\def \lc {light cone\ }
\def \vv {{\cal X}}
\def \X {{\cal X}}

\def\bea{\begin{eqnarray}}
\def\eea{\end{eqnarray}}
\def\beann{\begin{eqnarray*}}
\def\eeann{\end{eqnarray*}}
\def\beq{\begin{equation}}
\def\eeq{\end{equation}}
\def\ba{\begin{array}}
\def\ea{\end{array}}
\def\ben{\begin{enumerate}}
\def\een{\end{enumerate}}
 \def \l {\lambda}
 \def\m {\mu}
\def\s {\sigma }

 \def \la {\label}
 \def\be{\begin{equation}}
\def\ee{\end{equation}}

\def \ci {\cite}
\def \la {\label}
\def \const {{\rm const}}
\def \haf{{\textstyle { 1 \ov 2}} }
\def \de{{\textstyle { 1 \ov 9}} }
\def \si{{\textstyle { 1 \ov 6}} }
\def \rt {{\tx { \ta \ov 2}}}
\def \r {\rho}
\def \fo{{ 1 \ov 4}}

\def \he {{\rm h}}

\font\mybb=msbm10 at 11pt
\font\mybbb=msbm10 at 17pt
\def\bb#1{\hbox{\mybb#1}}
\def\bbb#1{\hbox{\mybbb#1}}
\def\bZ {\bb{Z}}
\def\bR {\bb{R}}
\def\bE {\bb{E}}
\def\bT {\bb{T}}
\def\bM {\bb{M}}
\def\bH {\bb{H}}
\def\bC {\bb{C}}
\def\bA {\bb{A}}
\def\e  {\epsilon}
\def\bbC {\bbb{C}}
\def \DD {{\rm D}}
\def \foot {\footnote}
\def \k {\kappa}
\def \ov {\over}
\def \ha { { 1\ov 2}}
\def \we { \wedge}
\def \P { \Phi} \def\ep {\epsilon}
\def \go { g_1}\def \gd { g_2}\def \gt { g_3}\def \gc { g_4}\def \gp { g_5}
\def \F {{\cal F}}
\def \del { \partial}
\def \t {\theta}
\def \p {\phi}
\def \ee {\epsilon}
\def \te {\tilde \epsilon}
\def \ps {\psi}
\def \td {\tilde}
\def \g {\gamma}
\def \bi{\bibitem}
\def\a{\alpha }
\def \p {\phi}
\def \ep {\epsilon}
\def \s {\sigma}
\def \gr {\rho}
\def \r {\rho}
\def \d {\delta}
\def \G {\Gamma}
\def \l {\lambda}
\def \m {\mu}
\def \g {\gamma}
\def \n {\nu}
\def \vp {\varphi}
\def \td {\tilde}
\def \x {{\rm x}}
\def \tv {{\rm v}}
\def \four {{1 \ov 4}}
\def \b {\beta}

\def\be{\begin{equation}}
\def\ee{\end{equation}}
\def \ci {\cite}
\def \bi {\bibitem}
\def \la{\label}
\def \f {{\rm f}}

\def \foot {\footnote}

\def \u {u }
\def \v  {v}
\def \t {\tau}
\def\aa{ {\cal A} }
\def \z {{\rm w}}
\def \B {\Sigma}
\def \ww {\td u}
\def \pp {{\rm p}}

\setcounter{section}{0}
\setcounter{subsection}{0}

\newsection{Introduction}

The dynamics of the massless superstring modes
of the various superstring theories
admit a description in terms of ten-dimensional
effective supergravities.
The latter include, in particular, an infinite tower of  $\alpha'$
corrections.
{}From the world-sheet perspective,
the  equations of the
target-space fields are
vanishing conditions
for the sigma model beta functions, i.e.
conditions for  conformal invariance.
In this approach, $\alpha'$ is the loop-counting parameter in the sigma-model
perturbation theory. At zeroth order in $\alpha'$, the vanishing of the
beta functions is equivalent to
the field equations of ordinary  supergravity theories.
Terms linear or higher in $\alpha'$ are associated with corrections
involving quadratic or higher polynomials
in the spacetime curvature and their supersymmetric completions.

{}For generic heterotic-string backgrounds
the $\alpha'$-corrections to sigma-model couplings are not known.
As a result, most configurations relevant to string
theory that have appeared in the literature,
like the  compactifications of \cite{Candelas, sw, wsd}, 
the five-brane \cite{Callan}
and various five-brane intersections \cite{Gauntlett, Papates}, are solutions
of ordinary supergravity theories.
In cases where there is sufficient world-volume supersymmetry, one
can argue that higher-order corrections are absent, see
 \cite{howe} for a general argument.
In general however, the investigation of stringy effects
requires taking the $\alpha'$ corrections into account. One such background
is the heterotic five-brane \cite{Callan} which emerges
at one loop in sigma-model perturbation. This and other related results 
have been extended to three loops in \cite{hpap}.

The low-energy dynamics of the heterotic string is
given by $N=1$ supergravity in ten dimensions.
The bosonic fields of the theory
are the spacetime metric $G$, the NS three-form field strength $H$,
the dilaton $\Phi$
and the gauge field $A$. The Green-Schwarz 
anomaly-cancellation mechanism  requires that the three-form Bianchi identity
receive an $\alpha'$ correction of the form
\be
dH=-{\alpha'\over 4} \left(p_1(M)-p_1(E)\right)+O(\alpha'^2),\nn
\label{oneone}
\ee
where $p_1(M)$, $p_1(E)$ are the first Pontrjagin forms  of
spacetime $M$ and of
the vector bundle $E$ with connection $A$, respectively.
In the $\alpha'$ expansion for $H$,
$ H=T+\alpha' f+ O(\alpha'^2)$,
the lowest-order term $T$ should be a {\it closed three-form}, i.e. $dT=0$.
This stems from the fact that at tree-level in the sigma model,
the string couples to a
two-form gauge potential $b$, where $T=db$.
On the other hand,
if $p_1(M)\not=p_1(E)$,
then $df\not=0$ and so $dH\not=0$.
Global anomaly cancellation requires in addition
that $dH$ be {\sl exact}, i.e. that $H$ be globally defined.

A class of heterotic-string backgrounds for which the  Bianchi identity
of the three-form $H$ receives
a correction of the type (\ref{oneone}) are those with (2,0) world-volume
supersymmetry. Such models were considered in  \cite{Hull}. The target-space
geometry of  (2,0)-supersymmetric sigma models
 has been extensively investigated in \cite{Hull, strom, howe}.
Recently,
there is  revived interest in these models \cite{lust, Waldram}
as string backgrounds and in connection to
heterotic-string compactifications with fluxes
\cite{beckera, beckerb}.

In this paper we investigate the $\alpha'$ corrections
to heterotic-string backgrounds with (2,0)-world-sheet
supersymmetry. We take spacetime to be
$M=\bR^{10-2n}\times X_n$ and demand that the background preserve
$2^{1-n}$ of  spacetime supersymmetry. 
The $n=2$ case was examined in \cite{hpap}.
The manifold
$X_n$ is Hermitian equipped with a compatible
connection $\nabla^{(+)}$ with skew-symmetric torsion $H$, i.e.
 $X_n$ is K\"ahler  with torsion (KT).
At first order in $\alpha'$, the holonomy of the connection
$\nabla^{(+)}$ is contained in $SU(n)$ and $X_n$ is  conformally balanced,
see appendix A.

We show that at linear order in $\alpha'$ 
such spacetime-supersymmetric backgrounds
satisfy the anomaly-cancellation
condition and the field equations.
In the proof, we make use of the results of \cite{gpsib}
summarised in appendix A of the present paper.
We find that the corrections to the other fields are determined
by the corrections to the metric. We stress
that {\sl consistency} of the anomaly cancellation condition with
the field equations requires that in
the latter we include the {\sl two-loop} contribution. A
sufficient condition for the consistency of 
spacetime supersymmetry with the anomaly cancellation
and the field equations, is the applicability
of the $\partial\bar\partial$-lemma\footnote{The $\partial\bar\partial$-lemma
is not  valid on all non-K\"ahler Hermitian manifolds.} on $X_n$.

We also consider the  Donaldson
equations on a non-K\"ahler Hermitian manifold. These
are related to the gaugino Killing-spinor equation.
{}For generic Hermitian manifolds, it is not apparent that 
the Donaldson equations
are associated with a second-order equation for the gauge connection,
i.e. a field equation. We find that they are, however,
 if the underlying Hermitian manifold
is conformally balanced.

It has been shown in \cite{gpsi} that
if $X_n$ is compact, nonsingular
and all fields are smooth, then $T$ vanishes and
to zeroth order in the $\alpha'$ expansion
$X_n$ is a Calabi-Yau $n$-fold.
We shall take this to be the starting point of
the $\alpha'$ expansion.
Proceeding to first order in $\alpha'$, $X_n$ is deformed
to a conformally-balanced KT manifold with
${\rm hol}(\nabla^{(+)})\subseteq SU(n)$.
We determine the deformations of the fields using Hodge theory.
Moreover, we compute the dimension of the moduli
space and we find it to be the same as that 
of the moduli
space of the underlying
Calabi-Yau manifold.  We therefore conclude that at this order in $\alpha'$,
 there is {\sl no lifting} of moduli in (2,0) compactifications with $T=0$.
As particular examples of the general theory mentioned above,
we compute the ${\cal O}(\alpha')$ corrections to the
 conifold \cite{co} and to the 
$U(n)$-invariant Calabi-Yau metric found in
 \cite{calabi}. We find that the
singularity of the conifold persists to first order in $\alpha'$.
For other $\alpha'$ corrections to conifold geometry see \cite{frolov}.

 This paper is organised as follows: In section two,
we establish our notation and
write down the field
and the Killing-spinor equations
for the heterotic string,
up to two loops in sigma-model perturbation theory 
(order ${\cal O}(\alpha')$).
In section three, we give  the conditions
 on the deformations of the metric required by spacetime supersymmetry,
  and express the deformations
 of the NS three-form and
the dilaton in terms of those of the metric.
 We then show that the Killing-spinor 
equations for the aforementioned set of fields
imply the field equations at this order
in $\alpha'$, provided the anomalous Bianchi identity of
the $H$ field is satisfied.
In section four, we show that the gaugino Killing-spinor
 equation, which is equivalent to
the Donaldson equations on a Hermitian
 manifold, implies the field equation for the gauge connection.
 In section five, we compute the ${\cal O}(\alpha')$ corrections
to the fields and show that the dimension
 of the moduli space  is the same as that
of the moduli space 
 of the Calabi-Yau space we started with at zeroth order in $\alpha'$.
 In section six and seven, we compute the ${\cal O}(\alpha')$ corrections
to the conifold and to the Calabi metric, respectively. In section
 eight, we discuss the consequences of our results in the context
 of compactifications with fluxes and we comment on $\alpha'$
 corrections beyond two-loops. In appendix A, we summarise some
 of the properties of KT geometry. In appendix B, we explain the
 relation between the Lichnerowicz and Laplace operators. Finally,
in appendix C
we give a solution to the
field equations at linear order in $\alpha'$ without
the use of the Killing-spinor equations.

\newsection{Field and Killing-spinor equations}

The bosonic fields of the ten-dimensional  supergravity
which arises as low energy effective theory of the heterotic
string  are the spacetime
metric $G$, the NS  three-form field strength $H$, the  dilaton $\Phi$
 and the gauge connection $A$.
We define the connections
$$
\nabla^{(\pm)}_M Y^N=\nabla_MY^N\pm {1\over2} H^N{}_{MR} Y^R~,
$$
where $\nabla$ is the Levi-Civita connection of the metric $G$ and 
$M,N, R=0,1\dots,9$
are spacetime indices.
The three form $H$ has an expansion in $\alpha'$ of the form\footnote{
Our form conventions are
$\omega_k={1\over k!} \omega_{i_1, \dots, i_k} dx^{i_1} \wedge\dots \wedge
dx^{i_k}$.}
\bea
H=T-{\alpha'\over4} \left( Q_3(\Gamma^{(-)})-Q_3(A)\right)+O(\alpha'^2)~,
\eea
where $T$ is a closed three-form, 
$dT=0$ and  $Q_3$ are Chern-Simons three-forms. We have
\beq
dH\equiv -\alpha' P+O(\alpha'^2)~,~~~~~~~~P={1\over4}
[{\rm tr} (R^{(-)}\wedge R^{(-)})-{\rm tr}( F\wedge F)]~,
\la{pontrj}
\eeq
where the trace on the gauge indices is taken as
$$
{\rm tr} F\wedge F= F^a{}_b\wedge F^b{}_a~,~~~~~~~~~F=dA+A^2~.
$$
Similarly for the trace of $R^{(-)}$, where $R^{(-)}$ is
the curvature of the connection $\nabla^{(-)}$.
The four-form $P$ is proportional to the difference
of the Pontrjagin forms of the tangent
bundle of spacetime and Yang-Mills bundle of the heterotic string.

The string frame field equations of the heterotic string up to two-loops \cite{hta}  in sigma model
perturbation theory are
\bea
R_{MN}+{1\over4} H^R{}_{ML} H^L{}_{NR}+2\nabla_M\partial_N\Phi~~~~~~~~~~~~&&
\cr
+ {\alpha'\over4}
[ R^{(-)}{}_{MPQR} R^{(-)}{}_N{}^{PQR}
- F_{MPab} F_N{}^{Pab}]
+O(\alpha'^2)&=&0
\nonumber\\
\nabla_M\big(e^{-2\Phi} H^M{}_{RL}\big)+O(\alpha'^2)&=&0
\nonumber\\
\nabla^{(+)}{}^M(e^{-2\Phi} F_{MN})+O(\alpha'^2)&=&0~,
\label{feq}
\eea
where  we have suppressed the
gauge indices.
The field equation  of the dilaton $\Phi$ is implied from the
 first two equations above.  Our curvature conventions
are given in appendix A.

Let $\{\Gamma^M; M=0,\dots,9\}$
be a basis of the Clifford algebra Cliff($\bR^{1,9}$), i.e.
$\Gamma^M\Gamma^N+\Gamma^N\Gamma^M
=2G^{MN}$.
Then the string frame Killing-spinor equations\footnote{We have  used the notation
$\Gamma^{M_1\dots M_k}=\Gamma^{[M_1}\dots \Gamma^{M_k]}$.}
 are \cite{strom, roo}
\bea
\nabla^{(+)} \epsilon+ O(\alpha'^2)&=&0
\nonumber\\
\big(\Gamma^M\partial_M\Phi-
{1\over12} H_{MNR} \Gamma^{MNR}\big)\epsilon+O(\alpha'^2)&=&0
\nonumber\\
F_{MN}\Gamma^{MN}\epsilon+O(\alpha'^2)&=&0~ ,
\label{keq}
\eea
where $\epsilon$ is a section of the
 spin bundle $S_+$.\footnote{
The spin group $Spin(1,9)$ has two
inequivalent irreducible sixteen-dimensional
spinor representations and $S_\pm$ are the associated bundles.} 
It is clear that the first Killing
 spinor equation is a parallel
transport equation for the connection $\nabla^{(+)}$.
Since the connection
of the spin bundle $S_+$ is induced from the
tangent bundle of spacetime,
the investigation of this Killing-spinor equation
is greatly simplified. The first, second
and third Killing-spinor equations are associated with
the supersymmetry transformations of the
gravitino, dilatino and gaugino, respectively.
We shall use this terminology in what
follows to distinguish between them.

It is clear from the field equations that the  two-loop contribution to
the Einstein equations 
is at the same order as the modification of the torsion $H$
due to the cancellation of the heterotic anomaly.  
Consistency then
requires 
that both should be taken into account. The various field and Killing-spinor
equations are expected to receive corrections to all orders in $\alpha'$.
Therefore a solution of the field and/or Killing-spinor equations
of the effective supergravity theory can be expanded as
\bea
G&=&g+\alpha' h+O(\alpha'^2)
\nonumber\\
H&=&T+\alpha' f +O(\alpha'^2)
\nonumber\\
\Phi&=&\varphi+\alpha'\phi+O(\alpha'^2)
\nonumber\\
A&=& B+\alpha' Q+O(\alpha'^2)~.
\label{expeq}
\eea
In this expansion, the fields $(g, T, \varphi, B)$ solve
the field and the Killing-spinor equations at  zeroth-order in $\alpha'$.
We again remark that $dT=0$ although $dH$ may {\it not} vanish, $dH\not=0$.
The deformation $(h, f, \phi, Q)$ linear in $\alpha'$ is the
first-order correction to the background $(g, T, \varphi, B)$.
Of course, the fields receive higher-order corrections in $\alpha'$.
In what follows, we  determine the deformations $(h, f, \phi, Q)$ by requiring that
$(G,H,\Phi, A)$ in (\ref{expeq}) solve the field (\ref{feq})
and Killing-spinor (\ref{keq})  equations.

\section{The $\alpha'$ corrections to backgrounds with torsion}

\subsection{ World-sheet and spacetime supersymmetry}

We  restrict our attention to heterotic string backgrounds
of the form
\bea
ds^2&=&ds^2(\bR^{10-2n})+ds^2(X_n)
\nonumber\\
T&=&{1\over3!} T_{ijk}(y) dy^i\wedge dy^j\wedge dy^k
\nonumber\\
\varphi&=&\varphi(y)
\nonumber\\
B&=&B_i(y) dy^i
\label{comanz}
\eea
where
 $\{y^i; i=1,\dots, 2n\}$ are coordinates on a
manifold  $X_n$,  $n\leq 4$, and $dT=0$ as we have explained in
the introduction.

In addition, we  require that the background $(g, T, \varphi, B)$ 
be compatible with
(2,0) world-sheet supersymmetry. This means that  the light-cone gauged fixed
string world-sheet action is (2,0)-supersymmetric.
In particular this implies that   $X_n$ is a hermitian manifold,
$(X_n, J, g)$, with complex structure
$J$ which is parallel with respect to $\nabla^{(+)}$ connection, i.e.
$(X_n, J, g)$ is a KT manifold. The torsion $T$ of KT manifolds is specified
by the metric $g$ and the complex structure $J$, see appendix A.

The  background (\ref{comanz}) is expected to receive $\alpha'$
corrections because the supergravity field equations are modified
by two- and higher-loop contributions   in sigma model perturbation
theory and in particular by the heterotic anomaly-cancellation mechanism.
After these  corrections are included,  the background
is expected to be of the form
\bea
ds^2&=&ds^2(\bR^{10-n})+d\tilde s^2(X_n)
\nonumber\\
H&=&{1\over3!} H_{ijk}(y) dy^i\wedge dy^j\wedge dy^k
\nonumber\\
\Phi&=&\Phi(y)
\nonumber\\
A&=&A_i(y) dy^i~,
\label{comanza}
\eea
where $d\tilde s^2(X_n)=G_{ij}(y) dy^i dy^j$.
The three-form $H$ is {\it not} necessarily closed,
 because of (\ref{pontrj}).

As we have seen,  sigma model loop effects and the heterotic anomaly
cancellation mechanism alter the geometry of the manifold $X_n$.
Nevertheless, it is expected that if the original manifold $(X_n, J, g)$ has
a KT structure, the  geometry, after the corrections are taken
into account, remains KT.
So the manifold $(X_n, J, G)$  has a KT  structure as well but now
the torsion $H$ is {\it not} closed.
This is because it is
expected that there is a scheme
which preserves the (2,0) world-volume  supersymmetry
 in sigma model perturbation theory \cite{howegpc}.

A solution $(g, T, \varphi, B)$ of the zeroth order in $\alpha'$ field equations
associated  with KT manifold $(X_n,J,g)$ does {\it not} necessarily satisfy
the Killing-spinor equations (\ref{keq}) of supergravity theory. The conditions
for $(g, T, \varphi, B)$ to satisfy the gravitino, dilatino and gaugino
Killing-spinor equations \cite{strom, gpsi}  are
\bea
{\rm hol}(\nabla^{(+)})\subseteq SU(n) ~&,&~~~~~~~~~~~~~~~~~\theta=2d\varphi
\cr
F(B)_{2,0}=F(B)_{0,2}=0~&,& ~~~~~~~~~~~~\Omega^{ij} F(B)_{ij}=0~,
\label{kcon}
\eea
where ${\rm hol}(\nabla^{(+)})$ is the
holonomy of the connection $\nabla^{(+)}$, $\Omega_{ij}=
g_{jk} J^k{}_j$ is the K\"ahler form
and  $\theta$ is the Lee form of the
 Hermitian geometry.  (The Lee-form has been given
in appendix A). KT manifolds for which the Lee-form is
 {\it exact} are called {\it conformally
balanced}. The conditions on the curvature $F(B)$ of the gauge connection
$B$ required by the gaugino Killing-spinor equations imply
that $F$ is a (1,1)-form with respect to the complex structure $J$
and its trace with $\Omega$ vanishes. 
I.e. considered as a two-form $F(B)$ takes values
 in the Lie algebra of $SU(n)$. These conditions 
are the analogue of the Donaldson
 equations for Hermitian manifolds.
 It can be shown that the backgrounds of (\ref{kcon}) 
preserve $2^{1-n}$ of spacetime supersymmetry.

Conversely if $(g, T, \varphi, B)$ in (\ref{comanz})
satisfies the Killing-spinor
equations (\ref{keq}) preserving $2^{1-n}$ of spacetime supersymmetry,
then $X_n$ is a conformally balanced KT manifold and
the holonomy of $\nabla^{(+)}$ is contained in $SU(n)$.

As we have explained, 
the geometry of the background $(G, H, \Phi, A)$ (\ref{comanza})
is expected to be KT. However, it is not apparent that if the
$(g, T, \varphi, B)$  background
is
 spacetime supersymmetric, then $(G, H, \Phi, A)$ will also 
be spacetime supersymmetric.
The
corrections to Killing-spinor equations
of supergravity (\ref{kcon})
at order ${\cal O}(\alpha')$ are
determined by the corrections
to the metric and
the torsion but otherwise their
dependence on the metric and the torsion remains the same
\footnote{It is not expected that this property of the Killing
spinor equations  persists to all 
orders in sigma model perturbation
theory. The dependence of the Killing-spinor equations on the metric 
and the torsion
will change at higher orders \cite{italians}.}.
This has the consequence that 
if we insist that the corrected background $(G,H, \Phi, A)$
preserve the same number of supersymmetries as $(g, T, \varphi, B)$, then
$(X_n, J, G)$ is again a KT manifold for which 
${\rm hol}\nabla^{(+)}\subseteq SU(n)$,
 $\theta=2d\Phi$ and $F(A)_{2,0}=F(A)_{0,2}=0$, $\Omega^{ij} F_{ij}=0$. 
In this case,
 $\nabla^{(+)}, \theta$ and $\Omega$ are the connection, Lee form and K\"ahler
 form of the metric $G$, respectively.

We conclude that at linear order in  $\alpha'$, the corrections to the
geometry of $X_n$  
are deformations which preserve the following two properties:
\begin{itemize}

\item{}$X_n$ is a conformally balanced KT manifold and

 \item{} the holonomy $\nabla^{(+)}$ is contained in $SU(n)$.

\end{itemize}

In what follows, we  derive the
conditions on the deformations of the geometry which preserve the above properties and we solve the field and Killing-spinor
equations to first order in $\alpha'$. We  also present a similar analysis
for the conditions on the gauge connection.

\subsection{Gravitino and dilatino Killing-spinor equations}

As we have mentioned,
 in order to solve the gravitino and dilatino Killing-spinor equations,
we have to specify the
deformations $(G,H)=(g+\alpha' h, T+\alpha' f)$  which preserve the
properties that  $(X_n, J, g)$ is conformally balanced KT manifold and 
 ${\rm hol}(\nabla^{(+)})\subseteq SU(n)$.

{}First, we  consider  deformations which preserve the
hermiticity of the metric with respect to the complex structure $J$. This
means that  $h_{\alpha\beta}=0$,
where $\alpha, \beta=1, \dots, n$ are labels for the
holomorphic coordinates on $X_n$.
It can then be easily shown that ${\rm hol}(\nabla^{(+)})\subseteq U(n)$
 provided that the deformation for the torsion is
\be
f_{\alpha\beta\bar\gamma}=-\nabla_\alpha h_{\beta\bar\gamma}
+\nabla_\beta h_{\alpha\bar\gamma}~,
\label{torf}
\ee
where $\nabla$ is the Levi-Civita connection of the metric $g$,
$f_{\bar\alpha\bar\beta\gamma}= (f_{\alpha\beta\bar\gamma})^*$ and the rest
of the components vanish. The latter is required because the torsion
of a KT geometry is a (2,1)- and (1,2)-form.

Furthermore, the deformation of the connection
of the canonical bundle\footnote{This is the diagonal
$U(1)$ part of a  $U(n)$ connection on the tangent bundle.} of $X_n$
associated with $\nabla^{(+)}$
is
\bea
\omega_\alpha(G)&=&\omega(g)_\alpha+\alpha' [2i \nabla_\beta h_\alpha{}^\beta-
i \nabla_\alpha h^\beta{}_\beta
\cr
&+&
i T_{\bar\delta}{}_{\beta\bar\gamma} g^{\beta\bar\gamma} h_{\alpha\bar\delta}-
iT_{\alpha\beta\bar\gamma} h^{\beta\bar\gamma}]+ O(\alpha'^2)
\label{cacon}
\eea
where $\omega(g)$ is the  connection of the canonical bundle
associated with the connection\footnote{To avoid confusion, sometimes we use the notation
$\nabla^{(+)}(g)$ to denote the connection $\nabla^{(+)}$ with respect
to the metric $g$.}
  $\nabla^{(+)}(g)$
 and $\omega_{\bar\alpha}=(\omega_{\alpha})^*$.
 A necessary and sufficient condition for ${\rm hol}(\nabla^{(+)})\subseteq SU(n)$
 is that the curvature of the canonical bundle vanishes.
{}For the connection (\ref{cacon}), a sufficient condition is
\be
2i \nabla_\beta h_\alpha{}^\beta-
i \nabla_\alpha h^\beta{}_\beta
+
i T_{\bar\delta}{}_{\beta\bar\gamma} g^{\beta\bar\gamma} h_{\alpha\bar\delta}-
iT_{\alpha\beta\bar\gamma} h^{\beta\bar\gamma}=0~,
\label{uone}
\ee
where $\nabla=\nabla(g)$.

It remains to find the condition required for $X_n$ to remain conformally
balanced after the deformation.
 {}For this, we compute the first-order deformation of the Lee form
to find
\be
\theta_\alpha=\theta(g)_\alpha+\alpha'
 [-\nabla_\alpha(g^{\beta\bar\gamma} h_{\beta\bar\gamma})
-\nabla_\beta h_{\alpha\bar\gamma} g^{\beta\bar\gamma}+{1\over2}
 T_{\alpha\beta\bar\gamma} h^{\beta\bar\gamma}-{1\over2}
 H^{\bar\delta}{}_{\beta\bar\gamma}  g^{\beta\bar\gamma} h_{\alpha\bar\delta}~,
 \label{theta}
 \ee
where $\theta(g)$ is the Lee-form of the $(g,J)$ geometry.
Substituting (\ref{uone}) in (\ref{theta}), we find that
\be
\theta_\alpha=\theta(g)_\alpha+
{\alpha'\over2} \nabla_\alpha(g^{\beta\bar\gamma} h_{\beta\bar\gamma})
\ee
The dilatino Killing-spinor equation can be solved by
setting
\be
\phi={1\over4} g^{\beta\bar\gamma} h_{\beta\bar\gamma}~.
\label{cordil}
\ee
Therefore the dilaton is deformed to
$$
\Phi=\varphi+{\alpha'\over4} g^{\beta\bar\gamma} h_{\beta\bar\gamma}+O(\alpha'^2)~.
$$

The  equations that remain to be solved are those in (\ref{uone}).
Observe that these equations are $2n$ in number, 
i.e. as many as the diffeomorphisms of
$X_n$. Since there is some redundancy in specifying the deformation
$h$ up to an infinitesimal  diffeomorphism generated
by the vector field $v$, i.e. $h'_{\alpha\bar\beta}=h_{\alpha\bar\beta}
+\nabla_\alpha v_{\bar\beta}+\nabla_{\bar\beta} v_{\alpha}$, it is expected
on physical grounds that it is always  possible to choose an $h$ so that (\ref{uone})
is satisfied. Therefore (\ref{uone}) can be viewed as   a choice of gauge fixing
for diffeomorphisms of $X_n$.
In the following,  we provide more evidence that this  is a good gauge choice.

It remains to examine the conditions on the gauge connection.
We postpone  this for after the investigation of the field
equations of the metric and the two-form gauge potential.

\subsection{The solutions of field equations}

Having derived the conditions for 
the deformations to
satisfy the gravitino and dilatino Killing-spinor
equations, we  now focus on the solutions of the field equations for the
metric and the NS two-form  potential. In particular we  show that at
order $\alpha'$ both these field equations are satisfied
 provided that the heterotic anomaly-cancellation
condition holds.

Assuming that the background $(g, T, \varphi)$ satisfies the field equations
at zeroth order in $\alpha'$, substituting   (\ref{expeq}) in the field equation
for the metric $(\ref{feq})$ and
collecting the terms linear in $\alpha'$, we find
\bea
\Delta_Lh_{ij}&-&{1\over4}T_{imn} f_j{}^{mn}-{1\over4} T_{jmn} f_i{}^{mn}
+{1\over 2} h^{mn} g^{kl} T_{imk} T_{jnl}
\cr
&+& 2\nabla_i\partial_j \phi
-g^{kl} (\nabla_i h_{jk}+\nabla_j h_{ik}- \nabla_kh_{ij}) \partial_l\phi_0+S_{ij}=0~,
\label{dfeqm}
\eea
where $\Delta_L$ is the Lichnerowicz operator
with respect to the metric $g$ (see appendix B) and
$$
S_{ij}={1\over4} [R^{(-)}{}_{iklm} R^{(-)}{}_j{}^{klm}-F_{ik ab} F_j{}^{kab}]
$$
is the two loop contribution to the beta function.
The curvature $R^{(-)}$ is with respect to  $(g,T)$ and $F=F(B)$.

Clearly, (\ref{dfeqm}) is rather involved. To proceed, we take the original
background $(g, T,$ 
$\varphi, B)$  to be spacetime supersymmetric in the
 sense described in the previous section.
In addition we assume that after the deformation the
background remains supersymmetric. As we have
seen this is equivalent to requiring that the  KT geometry $(X_n, J, G)$
be conformally balanced and ${\rm hol}(\nabla(G)^{(+)})\subseteq SU(n)$.
Substituting the deformation (\ref{expeq}) in
(\ref{ktfeq}) of appendix A and collecting the terms linear in $\alpha'$,
we find that
\bea
\Delta_Lh_{ij}&-&{1\over4}T_{imn} f_j{}^{mn}-{1\over4} T_{jmn} f_i{}^{mn}
+{1\over 2} h^{mn} g^{kl} T_{imk} T_{jnl}
\cr
&+& 2\nabla_i\partial_j \phi
-g^{kl} (\nabla_i h_{jk}+\nabla_j h_{ik}- \nabla_kh_{ij}) \partial_l\phi_0
={1\over4} J^k{}_i df_{kjmn} \Omega^{mn}
\label{bbc}
\eea
Observe that there is no explicit contribution of the metric deformation
 on the right-hand-side of  (\ref{ktfeq}). This is because at zeroth order the
torsion is closed, $dT=0$.
Using (\ref{dfeqm}) and (\ref{bbc}), we find that
\be
{1\over4} J^k{}_i df_{kjmn} \Omega^{mn}+S_{ij}=0~.
\label{fs}
\ee

Next consider the anomaly-cancellation condition (\ref{pontrj})
which to linear order in $\alpha'$ can be
written as
\be
df=-P~,
\label{fp}
\ee
where $P$ depends on $(g, T, B)$.
Since we have assumed that the background $(g, T, \varphi, B)$ is
supersymmetric,  $R^{(-)}$ and $F$ satisfy the
conditions
\bea
R^{(-)}_{mn}{}^i{}_j J^m{}_k J^n{}_l&=&R^{(-)}_{kl}{}^i{}_j~,~~~~~~\Omega^{mn}
R^{(-)}_{mn}{}^i{}_j=0
\cr
F_{mn}{}^a{}_bJ^m{}_k J^n{}_l&=&F_{kl}{}^a{}_b~, 
~~~~~~~\Omega^{mn}F_{mn}{}^a{}_b=0~.
\label{doldol}
\eea
These conditions on $R^{(-)}$ can be easily deduced from the fact that
the holonomy of $\nabla^{(+)}$ is contained in $SU(n)$ and
$R^{(-)}{}_{ij,kl}=R^{(+)}{}_{kl,ij}$ provided that the torsion is closed
($dT=0$). The conditions on $F$ can be deduced from the Killing-spinor
equations of the gaugino.
Contracting the anomaly-cancellation condition\footnote{This derivation
of (\ref{fs}) from (\ref{fp}) is sensitive to the relative numerical
coefficient of the two-loop contribution to the metric field equation and
that of the anomaly-cancellation condition.}
 (\ref{fp}) with the K\"ahler form $\Omega$ and  using (\ref{doldol}),
 it is easy to see that the  field equation for the metric (\ref{fs})
 is satisfied.

In order to solve the field equations for the metric, it is sufficient to
solve the anomaly-cancellation condition (\ref{fp}). Substituting (\ref{torf})
in (\ref{fp}), we find that
\be
P=-2i \partial \bar\partial Y
\label{ddb}
\ee
where $Y_{ij}= h_{ik} J^k{}_j$. The global anomaly-cancellation condition
requires that $P$ be exact. Since $P$ is an exact, real (2,2)-form, if the
$\partial\bar \partial$-lemma
is valid on $X_n$, then there is a real (1,1)-form $Y$ {\sl globally defined} on $X_n$ such that
(\ref{ddb}) is satisfied. On Hermitian manifolds 
which are {\sl not} K\"ahler,
the $\partial\bar \partial$-lemma is {\sl not} 
valid in general. Therefore a {\sl sufficient}
condition for the existence of spacetime supersymmetric solutions
in backgrounds with non-vanishing
torsion, $T\not=0$, 
is the {\sl validity} of the $\partial\bar \partial$-lemma.

The solution to the anomaly-cancellation condition (\ref{ddb}) is
not unique. Indeed, if $Y$ satisfies (\ref{ddb}), then
\be
Y'=Y+\partial \bar w+\bar\partial w
\label{gag}
\ee
where  $w$ is a (1,0)-form, is also a solution. This gauge freedom in $Y$
is equivalent to specifying the deformation  $h$ in the metric
up to an infinitesimal diffeomorphism. This can be easily seen by setting
$v=-iw$. Therefore having determined $Y$ from
(\ref{ddb}), we still have the gauge freedom to solve the supersymmetry condition
(\ref{uone}).

The solution of (\ref{ddb}) up to a gauge transformation
(\ref{gag}) is not unique. The classes of independent solutions
are described by the Aeppli group $V^{1,1}(X_n)$  defined
by
$$
V^{1,1}={{\rm Ker}(i\partial\bar\partial:
\Lambda^{1,1}(X_n)\rightarrow  \Lambda^{2,2}(X_n))
\over \partial \Lambda^{0,1}(X_n)+\bar\partial \Lambda^{1,0}(X_n)}~,
$$
see \cite{gut} for a related discussion.
The dimension of this group is the dimension of the moduli space
of solutions of (\ref{ddb}). However it is not apparent that all elements
of this group are associated with 
spacetime-supersymmetric deformations. The latter
 should in addition satisfy (\ref{uone}).

It remains to show that 
the field equations of the NS two-form gauge potential are  satisfied
as well. The proof of this is based on an identity shown in \cite{gpsib}
 (corollary 3.2) which
can be stated as follows: Let $(X_n, J, G)$ be a conformally balanced KT manifold
with torsion $H$, $dH\not=0$,
and  ${\rm hol}(\nabla^{(+)})\subseteq SU(n)$,
then
\be
\nabla^iH_{ijk}=\theta_i\, H^i{}_{jk}~.
\label{feqktb}
\ee
This statement is valid irrespectively  of whether or not $G$ is a small perturbation
of another metric $g$.

Both KT structures $(X_n, J,g)$ and $(X_n, G,J)$ satisfy
the aforementioned  conditions because they are supersymmetric.  Therefore
 both
satisfy (\ref{feqktb}) with their respective torsions and Lee forms.
Using (\ref{feqktb}) for the $\alpha'$ corrected  background $(G, H, \Phi, A)$, the
field equation (\ref{feq}) for the NS two-form
\be
-2\partial_i \Phi H^i{}_{jk}+\nabla^iH_{ijk}+O(\alpha'^2)=0
\label{nsn}
\ee
can be written as
$$
(\theta_i-2\partial_i \Phi) H^i{}_{jk}+O(\alpha'^2)=0
$$
which vanishes identically because  $(X_n, J, G)$ is conformally balanced.
Since the field equations for the background $(g, T, \varphi, B)$ are satisfied
by assumption and as we have shown  the field equations (\ref{nsn})  for
$(G, H, \Phi, A)$ are satisfied as well,
the part  of (\ref{nsn}) linear in $\alpha'$
vanishes identically.   Therefore the
field equations for the NS two-form gauge potential
are satisfied without additional conditions
on the deformation $h$ of the metric.

\section{The gauge field}

The main purpose 
of this section is to show that the Killing-spinor equations
of the gaugino imply the field equations of the gauge field before  
and after the
$\alpha'$ corrections are taken into account. The simplest way to show this
is by investigating the properties
of gauge fields on conformally balanced KT manifolds.

\subsection{Gauge Fields on KT conformally balanced manifolds}

We first describe
the well-known relation between the Donaldson equations and the field
equations of a gauge connection on a K\"ahler manifold.
Let $E$ be a vector bundle over a {\it K\"ahler} manifold $(M, J,G)$
 equipped with a connection $A$
with curvature $F$. If $A$ satisfies the Donaldson equations
\be
F_{0,2}=F_{2,0}=0~,~~~~~~~~\Omega^{ij} F_{ij}=0~,
\label{don}
\ee
then it is straightforward to show that  $A$ solves the field equations
\be
\nabla^i F_{ij}=0~.
\label{gfeq}
\ee
The proof makes use of the Jacobi identities for $F$.
Donaldson has shown that if $E$ is a stable bundle over a complex surface
 $M$,
then there is a unique connection which
solves (\ref{don}).

Next suppose   that $E$ is a vector bundle over a {\it non}-K\"ahler
conformally-balanced KT manifold
$(M, J,G)$ equipped with a connection $A$
with curvature $F$. 
Donaldson equations (\ref{don}) can be easily generalised
to  KT manifolds by allowing $\Omega$ to be the K\"ahler
form of the Hermitian metric $G$. We shall show that the Donaldson equations
imply the field equations
\be
\nabla^{(+)}{}^i(e^{-2\Phi} F(A)_{ij})=0~,
\label{nfeq}
\ee
where $\nabla^{(+)}$ is the connection of the KT structure $(M,J,G)$ with torsion $H$
and Lee form $\theta=2d\Phi$.
{}For this we choose complex coordinates with respect to
the complex structure $J$ and rewrite (\ref{nfeq})
as
$$
-2\nabla_\gamma G^{\gamma\bar\beta} F_{\bar\beta\alpha}
+G^{\gamma\bar\beta} \nabla_\gamma F_{\bar\beta\alpha}-
{1\over2} H^{\bar\delta}{}_{\gamma\bar\beta} F_{\bar\delta\alpha} G^{\gamma\bar\beta}
-{1\over2}H^\delta{}_{\gamma\alpha} F_{\bar\beta\delta} G^{\gamma\bar\beta}=0
$$
The Jacobi identities  imply that
$$
\nabla_\gamma F_{\bar\beta\alpha}+\nabla_\alpha F_{\gamma\bar\beta}=-
\nabla_{\bar\beta} F_{\alpha\gamma}
$$
where
$$
\nabla_{\bar\beta} F_{\alpha\gamma}={1\over2}
H^{\bar\delta}{}_{\bar\beta\alpha} F_{\bar\delta \gamma}
+{1\over2}H^{\bar\delta}{}_{\bar\beta\gamma} F_{\alpha\bar\delta}~,
$$
and $\nabla$ is the Levi-Civita connection of the metric $G$.
Collecting the various terms together, we find that
$$
(\theta_\gamma-2\partial_\gamma\Phi) G^{\gamma\bar\beta} F_{\bar\beta\alpha}=0
$$
which vanishes identically because $(M, J,G)$ is conformally balanced,
 $\theta=2d\Phi$.
Therefore equations (\ref{don}) together with the
Jacobi equations imply the field equations (\ref{nfeq}).

\subsection{The gauge field equations}

We shall use the result of the previous section to show that
the field equations of the gauge field are satisfied provided that
the Killing-spinor equations (\ref{keq}) are satisfied.

Assuming that the background $(g, T, \varphi, B)$ and its deformation
$(G, H, \Phi, A)$ satisfy the Killing-spinor equations (\ref{keq}), the
KT structures $(X_n, J, g)$ and $(X_n, J, G)$ are conformally balanced
and the holonomies 
of their $\nabla^{(+)}$ connections are contained in $SU(n)$.
In addition both gauge connections $B$ and its deformation $A$ satisfy
the conditions
(\ref{don})--the latter up to linear order in $\alpha'$.

Applying the theorem 
proven 
in the previous section, we conclude that the background
$(g, T, \varphi, B)$ and its deformation
$(G, H, \Phi, A)$ satisfy the field equations (\ref{nfeq})--the latter up to
linear order in $\alpha'$. Expanding (\ref{nfeq}) in  $\alpha'$  for the background
$(G, H, \Phi, A)$ and since (\ref{nfeq}) is satisfied
at  zeroth order in $\alpha'$,
the linear term in $\alpha'$ will vanish identically. Thus
if the background $(g, T, \varphi, B)$ and its deformation
$(G, H, \Phi, A)$ satisfy the Killing-spinor equations, then the field equations
(\ref{feq}) for the gauge potential
are satisfied without additional conditions  on the deformations. The
conditions on the deformation $Q$ of the gauge connection $B$ are
\bea
\nabla_\alpha Q_\beta-\nabla_\beta Q_\alpha =\nabla_{\bar\alpha} Q_{\bar\beta}-
\nabla_{\bar\beta} Q_{\bar\alpha}&=&0
\cr
h^{\alpha\bar\beta} F(B)_{\alpha\bar\beta}+ g^{\alpha\bar\beta}
\left(\nabla_\alpha Q_{\bar\beta}-\nabla_{\bar\beta} Q_\alpha\right) &=&0~,
\label{gaugesu}
\eea
where the covariant derivative $\nabla$ is with respect to the gauge connection $B$.
We have derived these by substituting the deformation $(G, H, \Phi, A)$ of
$(g, T, \varphi, B)$ in
the Killing-spinor equations (\ref{don}) and by collecting the
 terms linear in $\alpha'$.

It remains to find whether the Killing-spinor equations for the gaugino or
equivalently (\ref{don}) have solutions on a general conformally
balanced KT manifold. It is easy to investigate the case where $A$
is an abelian connection. However, the non-abelian case is more involved and
so we shall not pursue this further.

\section{(2,0) heterotic compactifications}

The compactification ans\"{a}tze for the heterotic string which preserve
(2,0) world-volume supersymmetry and are 
spacetime supersymmetric, are given in (\ref{comanz})
with the additional assumption that the internal space $X_n$ is compact
\footnote{The ansatz (\ref{comanz}) was considered in \cite{strom} where
it was shown that no warp factor is allowed for the non-compact
part of the metric.  
Therefore, as is easy to see using sigma-model perturbation theory, 
there can be no such warp factor in (\ref{comanza}) either.}.
As we have discussed, such backgrounds are expected to
receive $\alpha'$ corrections. Using the machinery  developed
in the previous sections, we shall investigate the deformations
of these backgrounds due to $\alpha'$ corrections.

\subsection{The zeroth-order solution}

As we have explained the requirement for a compactification to
preserve (2,0) world-sheet supersymmetry and $2^{1-n}$ of
spacetime supersymmetry in $(10-2n)$-dimensions leads to an internal
manifold $X_n$ with a conformally balanced KT structure and
${\rm hol}(\nabla^{(+)})\subseteq SU(n)$.

The additional  requirement that $X_n$ is compact\footnote{This assumption
is sufficient for the spectrum in $(10-2n)$ dimensions to be discrete.},
the implicit assumption that all the fields $(g, T, \varphi)$ are smooth on $X_n$
and the fact that at zeroth order in $\alpha'$ $dT=0$, impose strong restrictions
on the geometry of $X_n$.
It has been shown in \cite{gpsi} under the above assumptions\footnote{
In fact in \cite{gpsi} it was shown that $X_n$ is Calabi-Yau
even if $dT\not=0$ provided a certain inequality holds.}
 that $X_n$ {\sl is} Calabi-Yau , $T=0$ and the dilaton $\varphi$ is constant.
In addition at zeroth order in $\alpha'$ the gauge connection $B$
satisfies the Donaldson equations (\ref{don}) on the Calabi-Yau manifold $X_n$.
These data are the starting point of our $\alpha'$ expansion.

The Calabi-Yau background $(g, \varphi, B)$
receives $\alpha'$ corrections. The deformation 
$(G,H,$ $\Phi, A)$ of
 $(g, \varphi, B)$ has non-vanishing torsion $H$
which is not closed, $dH\not=0$, as required by the
 anomaly-cancellation mechanism.
Note that the zeroth-order term in $H$ vanishes and so  $H$ is purely 
first-order in $\alpha'$,
$H=\alpha' f+O(\alpha'^2)$.

\subsection{The first-order solution}

To find the 
correction $(G, H, \Phi, A)$ to the Calabi-Yau geometry $(g, \varphi, B)$, we
 have to solve equations (\ref{uone}), (\ref{gaugesu})  for the  first-order
deformations $(h, f, \phi, Q)$.
The deformation $\phi$ to the dilaton is given in (\ref{cordil}),
$\phi={1\over4} g^{\beta\bar\gamma} h_{\beta\bar\gamma}$. Similarly,
the deformation $f$ of the torsion is given in (\ref{torf}).
The remaining field
and Killing-spinor equations are satisfied without further conditions.

Since for this background the torsion vanishes
at zeroth order in $\alpha'$,  condition
(\ref{uone})
arising from  the requirement that ${\rm hol}(\nabla^{(+)})\subset SU(n)$, can be rewritten
as
\be
\nabla^{\bar\beta} h_{\alpha\bar\beta}-{1\over2} \nabla_\alpha (g^{\gamma\bar\beta}
h_{\gamma\bar \beta})=0~.
\label{gaugfix}
\ee
The above equation can be thought of as a gauge-fixing condition
for the deformations associated with infinitesimal diffeomorphisms
of the underlying manifold.
Condition (\ref{gaugfix}) can always be attained.
To see this, first write (\ref{gaugfix}) in real coordinates
\be
\nabla^j h_{ji}-{1\over4} \nabla_i h^j{}_j=0~.
\label{gaugfixa}
\ee
Suppose that $h$ does not solve (\ref{gaugfixa}). We shall show  that there is
a $v$ such that $h'$ given by
  $h'_{ij}=h_{ij}+\nabla_i v_j+\nabla_j v_i$ satisfies
 (\ref{gaugfixa}). $v$ is determined by
\be
\nabla^k \nabla_k v_i+{1\over2}\nabla_i \nabla^k v_k
=\nabla^j h_{ji}-{1\over4} \nabla_i h^j{}_j~.
\label{ddf}
\ee
First note that the Kernel of the operator on the left-hand-side
of the equation above is zero on an
{\sl irreducible} Calabi-Yau manifold. Indeed let $v$ be in
the Kernel. Then
$$
\int_{X_n}( v^i \nabla^k \nabla_k v_i+{1\over2}v^i \nabla_i \nabla^k v_k)~ d{\rm vol}=-
\int_M \left(\nabla_k v_i \nabla^k v^i+{1\over2} (\nabla^k v_k)^2 \right) d{\rm vol} =0~.
$$
Thus $v$ is parallel with respect to the Levi-Civita
connection. Since $X_n$ is irreducible, there are no parallel one-forms on $X_n$
and so the Kernel vanishes.  Therefore equation  (\ref{ddf}) can be
solved for $v$ since the right-hand-side does not vanish, by assumption.

We can also determine the metric moduli of (2,0)
compactifications. Since the $\partial\bar\partial$-lemma applies
for Calabi-Yau manifolds, the Aeppli group $V^{1,1}(X_n)$ is
isomorphic to the the Hodge group $H^{1,1}(X_n)$.
 Thus the dimension of the  moduli space is the Hodge number
$h^{1,1}$ which is the same as the number of  metric moduli of
Calabi-Yau manifolds. Of course one can also take into account the moduli
associated with including in the theory NS two-form gauge
potentials. The latter are harmonic (1,1)-forms. This leads to a
complex moduli space of real dimension $2 h^{1,1}$. One
concludes that the metric moduli of  Calabi-Yau
(2,0)-compactifications, which are associated  with NS fluxes induced by the
heterotic anomaly, are {\sl not} lifted. An effect of the 
heterotic anomaly-cancellation mechanism
is a shift in the origin of the moduli space.

The moduli of the theory associated with deformations of the complex structure
of the underlying Calabi-Yau manifold is not lifted either. The analysis that
we have done using the complex structure $J$ can be repeated with any other
complex structure on the Calabi-Yau manifold. Therefore, we conclude
that the presence of NS flux associated with the heterotic anomaly
cancellation mechanism does {\sl not} lift the Calabi-Yau moduli at this
order in $\alpha'$ perturbation theory.

\section{The $\alpha'$-corrected conifold}

The conifold is a singular non-compact Calabi-Yau threefold \cite{co}.
Here we  apply
the machinery of the previous sections to compute
the $\alpha'$ corrections explicitly.

The conifold can  be thought of as a Ricci-flat
cone over a $T^{1,1}$ space, where the latter
is a particular $U(1)$ fibration over $S^2\times S^2$ \cite{pp}.
Let $0\leq \phi_{i}\leq 2\pi,~~0\leq \theta_{i}\leq \pi, ~~i=1,2$ be  angular
coordinates
parametrising the two spheres $S^2$, $0\leq \psi\leq 4\pi$
be the coordinate on the $U(1)$ fibre and $\rho\geq 0$ be
the radial coordinate. The line element of the conifold is
\bea
ds^2=g_{mn}dx^m dx^n&=&d\r^2
+{\r^2\over 9}(d\psi+cos\theta_1 d\p_1+cos\theta_2 d\p_2)^2\nn\\
&+&{\r^2\over6}(sin^2\theta_1 d\p_1^2+d\theta_1^2)
+{\r^2\over6}(sin^2\theta_2 d\p_2^2+d\theta_2^2)~.
\label{cnfld}
\eea
The second term on the right-hand-side is the vertical displacement along the
$U(1)$ fibre whereas the last two terms represent the line element of
$S^2\times S^2$. There is a conical singularity at $\r=0$,
where the $T^{1,1}$ base of the cone shrinks to zero size.
It is useful to note that
$T^{1,1}$ is topologically $S^2\times S^3$.

Let us now
consider a deformation $g_{mn}\rightarrow g_{mn}+\alpha' h_{mn}$,
where $h_{mn}$ is hermitian,
\be
h_{kl}J^{\kappa}{}_{m}J^{l}{}_{n}=h_{mn}.
\label{hermi}
\ee
The complex structure $J$ of the conifold is
\bea
J^{\theta_1}{}_{\p_1}=sin\theta_1&,&~~~~
J^{\r}{}_{\p_1}=-{\r\over 3}cos\theta_1,\nn\\
J^{\theta_2}{}_{\p_2}=sin\theta_2&,&~~~~
J^{\r}{}_{\p_2}=-{\r\over 3}cos\theta_2,\nn\\
J^{\p_1}{}_{\theta_1}=-{1\over sin\theta_1}&,&~~~~
J^{\psi}{}_{\theta_1}=cot\theta_1,\nn\\
J^{\p_2}{}_{\theta_2}=-{1\over sin\theta_2}&,&~~~~
J^{\psi}{}_{\theta_2}=cot\theta_2,\nn\\
J^{\r}{}_{\psi}=-{\r\over 3}&,&~~~~
J^{\psi}{}_{\r}={3\over \r}~,\nn
\eea
and the rest of the components vanish.
Condition (\ref{hermi}) implies that $h_{mn}$ is of the form,
\bea
h_{mn}dx^m dx^n&=&D [d\r^2
+{\r^2\over 9}(d\psi+cos\theta_1 d\p_1+cos\theta_2 d\p_2)^2]\nn\\
&+&A(sin^2\theta_1 d\p_1^2+d\theta_1^2)
+C(sin^2\theta_2 d\p_2^2+d\theta_2^2)\nn\\
&+&2B sin\psi(sin\theta_1 d\p_1 d\theta_2-sin\theta_2 d\p_2 d\theta_1)\nn\\
&+&2B cos\psi(d\theta_1d\theta_2+sin\theta_1 sin\theta_2 d\p_1 d\p_2)~.
\label{h}
\eea
We  take $A,~B,~C,~D$ to be functions of the
radial coordinate alone.
With this assumption,
it can be seen from the form of most general $T^{1,1}$ metric
given in \cite{mt}, that (\ref{h})
is a foliation of $T^{1,1}$ spaces.

The gauge-fixing condition (\ref{gaugfixa}) is equivalent to
\bea
0&=&A+C+{3\over 2}\r (A+C)'-{1\over 12}\r^3-{2\over 3}\r^2 F\nn\\
0&=&B
\label{gafi}
\eea
In order to solve the Einstein equation, we need to find
$$
S_{mn}:={1\over 4}R_m{}^{kls}R_{nkls}.
$$
After some computation, we get
\beq
S_{\p_1\p_1}=S_{\p_2\p_2}={sin^2\theta\over \r^2},~~
S_{\theta_1\theta_1}=S_{\theta_2\theta_2}={1\over \r^2}.\nn
\eeq
The Einstein equation (\ref{dfeqm}) 
(or its gauged-fixed version (C.4) of appendix C) 
is then equivalent to the following
set of conditions,
\bea
0&=&B\nn\\
0&=&A+C+{\r^2\over 24}(-8D+5\r D'+\r^2D'')\nn\\
0&=&6\r A'+6\r^2 A''+\r^2(-4D+5\r D'+\r^2D'')-3\nn\\
0&=&A'+\r A''-C'-\r C''
\eea
Demanding that the total metric be asymptotic to
the conifold in the region
$\r/\sqrt{\alpha'}\rightarrow \infty$,
the general solution to order ${\cal O}(\alpha')$ reads
\bea
A&=&-{1\over 16}
({3\over 2}-c_1-c_2~log{\rho\over \rho_0 })\nn\\
B&=&0\nn\\
C&=&-{1\over 16}
({3\over 2}+c_1+c_2~log{\rho\over \rho_0})\nn\\
D&=&-{3\over 8\r^2}
\label{einsol}
\eea
where $\rho_0$ is a dimensionful constant introduced to make
the constants  $c_1,c_2$  dimensionless.
Note that the gauge-fixing condition (\ref{gafi}) is
automatically satisfied by the solution (\ref{einsol}). Therefore
the solution
is spacetime supersymmetric.

To summarise, the ${\cal O}(\alpha')$-perturbed metric is
\bea
ds^2&=&(1-{3\alpha'\over 8\r^2}) [d\r^2
+{\r^2\over 9}
(d\psi+cos\theta_1 d\p_1+cos\theta_2 d\p_2)^2]\nn\\
&+&{\r^2\over 6}[1-{3\alpha'\over 8\r^2}
({3\over 2}-c_1-c_2~log{\rho\over \rho_0})]
(sin^2\theta_1 d\p_1^2+d\theta_1^2)\nn\\
&+&{\r^2\over 6}[1-{3\alpha'\over 8\r^2}
({3\over 2}+c_1+c_2~log{\rho\over \rho_0})]
(sin^2\theta_2 d\p_2^2+d\theta_2^2).
\label{pert}
\eea
The perturbation is valid in the regime $\r/\sqrt{\alpha'}>>1$. In
order to be able to extract any information about the behavior of the metric
near the apex ($\r=0$),
the full tower of $\alpha'$ corrections would have to
be taken into account as these become important in the sub-stringy
regime $\r/\sqrt{\alpha'} \leq 1$. This is clearly beyond the validity of
our analysis. Nevertheless, we can attempt to extrapolate our result to the
region $\r/\sqrt{\alpha'} \sim 1$. 
An analysis then reveals that at a radial distance of the order of $\sqrt{\alpha'}$,
an $S^2$ in the $T^{1,1}$ base
shrinks to zero volume.
We therefore find that the
singularity of the conifold persists, although it becomes milder in
the sense that not the entire base shrinks to zero.


\section{The $\alpha'$-corrected  $U(n)$-invariant Calabi-Yau metric}

To describe the $U(n)$-invariant Calabi-Yau metric \cite{calabi}, we
introduce  complex coordinates $\{z^\alpha; \alpha=1,\dots,n\}$ and
consider the $U(n)$ invariant Hermitian metric
$$
ds^2= A(r^2) dz\cdot d\bar z+B(r^2) \bar z\cdot dz~ z\cdot d\bar z~,
$$
where $r^2=\delta_{\alpha\bar\beta} z^\alpha  z^{\bar \beta}$, $dz \cdot d\bar z=
\delta_{\alpha\bar\beta} dz^\alpha dz^{\bar \beta}$ and similarly for the rest.
The metric is K\"ahler if
$$
B=A'
$$
where the prime denotes differentiation with respect to $r^2$.
The Levi-Civita connection one-form is
$$
\Gamma^\alpha{}_\beta= A^{-1} A'( \delta^\alpha{}_\beta \bar z\cdot dz
+ \bar z_\beta dz^\alpha)+{A''-2A^{-1} (A')^2\over A+r^2 A'}
 z^\alpha \bar z_\beta \bar z\cdot dz
$$
The holonomy of the above connection is contained $U(n)$.
The metric is Calabi-Yau if and only if the holonomy of the above connection
is in $SU(n)$. The connection of canonical bundle is
$$
\omega_\alpha=i \partial_\alpha {\rm ln} \det(g_{\alpha \bar\beta})~,
$$
where $\det(g_{\alpha\bar\beta})= A^{n-1} (A+r^2 A')$. The curvature
of canonical bundle vanishes iff
$$
A^n+ {r^2\over n} (A^n)'=\lambda~,
$$
where $\lambda$ is a constant. The most general solution of this equation is
$$
A^n=\lambda+{c\over r^{2n}}~,
$$
where $c$ is constant. This metric is the Calabi-Yau metric
on the orbifold $\bC^n/\bZ_n$ after resolving the singularity
at the origin by replacing with a $\bC P^2$.
Next we compute $P={1\over4} {\rm tr} R^2$ to find
$$
P={1\over4} C (dz\wedge d\bar z)\wedge (dz\wedge d\bar z)+
   {1\over4} C'  \bar z\cdot dz\wedge z\cdot d\bar z\wedge (dz\wedge d\bar z)~,
$$
where
$$
C=(n+1) A^{-2} (A')^2+ 2 A^{-1} A'  r^2 {A''-2A^{-1} (A')^2\over A+r^2 A'}
+ r^4 \left({A''-2A^{-1} (A')^2\over A+r^2 A'}\right)^2~
$$
and $(dz\wedge d\bar z)=\delta_{\alpha\bar\beta}~ dz^\alpha\wedge dz^{\bar\beta}$.
After  some computation, we find that
$$
C=n(n+1) \left({c\over \lambda r^{2n+2}+c r^2}\right)^2
$$
In addition setting
$$
h=D(r^2) dz\cdot d\bar z+ E(r^2)   \bar z\cdot dz~ z\cdot d\bar z~,
$$
and $Y_{\alpha\bar\beta}=-i h_{\alpha\bar\beta}$,
 we find that  $P=-2i \partial\bar\partial Y$ implies
\be
D'-E=-{1\over8}C~.
\label{defa}
\ee
Condition (\ref{gaugfix}) for $h$, gives
\be
2 (n-1) A^{-1} E- (n-1) A^{-1}D'-(n-1) A^{-2} A' D +{1\over \lambda}
[A^{n-1} (D+r^2 E)]'=0
\label{defb}~.
\ee
The supersymmetric deformation can be computed by solving (\ref{defa})
and (\ref{defb}).
Substituting for $E$ in (\ref{defb}), we find after some algebra that
\bea
0=&-&{c^2 n(n+1)\over 8\lambda r^{4n+4}}{3\lambda
+{nc\over r^{2n}}\over (\lambda+{c\over r^{2n}})^2}
-{c^2(n-1)\over \lambda r^{4n+2}}{1\over (\lambda+{c\over r^{2n}})}D\nn\\
&+& [{n-1\over \lambda}(\lambda-{c\over r^{2n}})
+2(\lambda+{c\over r^{2n}})]D'
+{r^{2}\over \lambda}(\lambda+{c\over r^{2n}})D''~.
\eea
We have not been able to obtain a closed form for $D$.
But we can solve the equation perturbatively in a large
distance expansion. The result reads,
\bea
E&=&{n(c/\lambda)^2\over 8 r^{4n+4}}\big(
n-2+{-5n^2+8n+1\over 2n+1} {(c/\lambda)\over r^{2n}}+\dots
\big)\nn\\
D&=&{n(c/\lambda)^2\over 8 r^{4n+2}}\big(
{3\over 2n+1}+{n^2-14n-3\over (2n+1)(3n+1)}
{(c/\lambda)\over r^{2n}}+\dots
\big)
\eea
Note that the leading correction to the metric behaves like $r^{-4n-2}$.

\section{Concluding Remarks}

Compactifications with fluxes lead to lower-dimensional
effective theories which exhibit potentials
lifting some of the moduli.
The relevant flux for (2,0) 
heterotic-string compactifications is the NS three-form $H$.
There are two possibilities. One is to
 allow for a non-vanishing flux at zeroth order in $\alpha'$.
Then the internal manifold is either non-compact
or/and some of the fields are singular. We have extensively investigated
the case of Calabi-Yau compactifications where the NS
flux vanishes at zeroth order in $\alpha'$.
As we have shown such compactifications develop  a non-vanishing
flux $H$  and a non-constant dilaton at
  first order  in $\alpha'$.
In addition, at this order in $\alpha'$,
 the compactifications have the same dimension of moduli space 
as that of the moduli space of 
the Calabi-Yau manifolds at zeroth order in $\alpha'$
 and so there is no lifting of moduli.

The effective lower-dimensional theories  which arise in standard
Calabi-Yau compactifications, i.e. without NS flux,
 may be different from those
with flux. If there is no NS flux 
induced by 
the heterotic anomaly-cancellation mechanism, then it is consistent
to take the lower-dimensional effective action to be 
the standard ten-dimensional N=1 supergravity action. Of course there
will be higher-order $\alpha'$ corrections. Nevertheless  it is consistent to 
consider
only the zeroth order in $\alpha'$. 
On the other hand, if one wishes to consider
the NS fluxes
induced by the heterotic anomaly-cancellation mechanism,
{\sl consistency requires} that curvature square terms in the field equations
and their supersymmetric completion should be considered as well.
These will contribute to the lower-dimensional effective theories.

The other possibility is  to introduce
a non-vanishing $H$ flux at zeroth order in $\alpha'$
and to allow for some fields to be singular
or/and for the internal manifold to be non-compact.
In the case that the  manifold or the fields are singular, one may expect that
these singularities can be resolved by taking 
into account $\alpha'$ corrections.
We have seen, for example, that after taking into account the linear order
$\alpha'$ correction to the conifold, the metric is less singular but the
singularity is not completely resolved.  In a scenario where the
singularities are removed and the internal manifold remains compact,
some moduli may be lifted and  potentials of the type given in  
\cite{gukov} may be generated
in the effective theory
in lower dimensions.

Similar conclusions can be drawn for compactifications of other theories 
with fluxes
in the presence of an anomaly-cancellation mechanism. Such an example
is M-theory \cite{dlm}. To investigate the consistency of the 
anomaly cancellation
with spacetime supersymmetry as we have done for the heterotic string, one
needs to know the higher-order derivative corrections to  $D=11$
supergravity and their supersymmetric completions.

It is of interest to speculate on
the geometry of the background (\ref{comanz})
after all $\alpha'$ corrections are taken into account.
It is expected that such a background is
a Hermitian manifold, based on 
the assumption that (2,0) world-sheet supersymmetry
is preserved in sigma-model perturbation theory.
It is unlikely that the 
supergravity Killing-spinor equations will remain of the
form  (\ref{keq}). 
The dependence on the metric and the torsion will change and
higher-curvature terms are expected to appear \cite{italians}.
So the conditions for preserving spacetime
supersymmetry will not be simply expressed as  conditions on the holonomy
of the $\nabla^{(+)}$ connection 
and on the Lee form $\theta$ of the Hermitian geometry.
However, some properties of the underlying manifold  may be preserved.
It is plausible to assume that the manifold  satisfies 
the $\partial\bar\partial$-lemma
and the canonical bundle is holomorphically trivial. It has been shown
in \cite{gut} that Moishezon manifolds with these properties admit a 
connection
with skew-symmetric torsion  and holonomy contained in $SU(n)$.
However, it is not apparent that the associated metric is that
which arises after taking into account all $\alpha'$ corrections. In addition,
on such non-K\"ahler Moishezon manifolds 
there is no KT structure  for which the associated  three-form
field strength is closed \cite{gut}. Therefore such manifolds 
cannot be used as the starting point of 
the sigma-model perturbation theory. 

\section*{Acknowledgements}

We would like to thank P. Howe for stimulating discussions. This work 
was partially supported by PPARC grants 
PPA/G/S/1998/00613 and PPA/G/O/2000/00451 and by EU grant HPRN-2000-00122.

\setcounter{section}{0}
\setcounter{subsection}{0}


\appendix{Useful Formulae for KT Geometry}

Let $(X_n, J, G)$ be a KT manifold, i.e. $X_n$ is a hermitian manifold of
complex dimension $n$
with metric $G$ and complex structure $J$ such that $\nabla^{(+)}J=0$, where
$\nabla^{(+)}$ has skew-symmetric torsion $H$.  In the mathematics literature
$\nabla^{(+)}$ is called the Bismut connection.
 The holonomy of  $\nabla^{(+)}$
is contained in $U(n)$. In complex coordinates, the holonomy
condition requires $\Gamma^{(+)}_i{}^\alpha{}_{\bar \beta}=0$ which in
turn gives
\be
H_{\alpha\beta\bar\gamma}=-\partial_\alpha G_{\beta\bar\gamma}
+\partial_\beta G_{\alpha\bar\gamma}~.
\label{ktto}
\ee
The rest of the components of the torsion are determined by complex conjugation.
The (3,0) and (0,3) components of $H$ vanish as it can be seen
from the integrability of the complex structure.
 So $H$ is determined
uniquely in terms of the metric and complex structure of $(X_n, J, G)$.
The Lee form of the KT geometry is
$$
\theta_i={1\over2} J^j{}_i H_{jkl} \Omega^{kl}~,
$$
where $\Omega_{ij}= G_{ik} J^k{}_j$ is the K\"ahler form.
In complex coordinates, the Lee form can be written as
\be
\theta_\alpha=\partial_\alpha{\rm ln}\det(G_{\beta\bar\gamma})- G^{\beta\bar\gamma}
\partial_\beta G_{\alpha\bar\gamma}
\label{ktth}
\ee
and $\theta_{\bar\alpha}= (\theta_\alpha)^*$.

The connection of the canonical bundle
induced by $\nabla^{(+)}$   is
\be
\omega_\alpha= i\Gamma^{(+)}{}_\alpha{}^\beta{}_\beta-
i\Gamma^{(+)}{}_\alpha{}^{\bar\beta}{}_{\bar\beta}=2i G^{\beta\bar\gamma}
\partial_\beta G_{\alpha\bar\gamma}-i G^{\beta\bar\gamma}\partial_\alpha G_{\beta\bar\gamma}
\label{ktuo}
\ee
and $\omega_{\bar\alpha}=(\omega_\alpha)^*$. Let $\rho=d\omega$ be the
curvature of the 
$U(1)$ connection $\omega$. The holonomy of the connection $\nabla^{(+)}$
is contained in $SU(n)$, iff $\rho=0$.

A KT manifold is conformally balanced iff there is a function $\Phi$ on $X_n$
such that $\theta=2d\Phi$, i.e. the Lee form is exact.
It has been shown in \cite{gpsib} that if
$(X_n, J,G)$ is a conformally balanced KT manifold and
the holonomy of $\nabla^{(+)}$
is contained in $SU(n)$ ($\rho=0$), then
\be
R_{ij}+{1\over4} H^k{}_{il} H^l{}_{jk}
+2\nabla_i\partial_j\Phi= {1\over4} J^k{}_i (dH)_{kjmn} \Omega^{mn}~,
\label{ktfeq}
\ee
where $\nabla$ is the Levi-Civita connection of the metric $G$. Note that
we do  {\sl not} require $dH=0$ in the above expression.

Our conventions for the curvature
of a connection $\Gamma$ are
$$
R_{ij}{}^k{}_l=\partial_i \Gamma^k_{jl}-\partial_j \Gamma^k_{il}+\Gamma^k_{im} \Gamma^m_{jl}-
\Gamma^k_{jm} \Gamma^m_{il}
$$
and the Ricci tensor is $R_{ij}=R_{mi}{}^m{}_j$.

\appendix{Lichnerowicz and Laplace operators}

Let $(M,g)$ be a Riemannian manifold with associated Levi-Civita connection $\nabla$.
The Lichnerowicz operator $\Delta_L$ is defined by
$$
R_{ij}(g+\epsilon h)= R_{ij}+\epsilon \Delta_L h_{ij}
+O(\epsilon^2)~.
$$
So $\Delta_L$ is the first-order deformation of the Ricci tensor.
A calculation reveals that
\bea
\Delta_Lh_{ij}=&-&{1\over2}\nabla^2 h_{ij}- R_{ik jl} h^{kl}
+{1\over2}\nabla_i\nabla^kh_{kj}+
{1\over2}\nabla_j\nabla^kh_{ki}
\cr
&-&{1\over2}\nabla_i\nabla_j h^k{}_k+{1\over2} R_{ki} h^k{}_j+{1\over2} R_{kj} h^k{}_i
\label{lop}
\eea
The Laplacian operator $\Delta$ on a two form $Y$ is
$$
\Delta Y_{ij}=-{1\over2} \nabla^k \nabla_k Y_{ij}-R_{ikj\ell} Y^{k\ell}+{1\over2}
R_{ik} Y^k{}_j-{1\over2}R_{jk} Y^k{}_i~.
$$
On Ricci-flat K\"ahler manifolds, we can relate the  Lichnerowicz operator
 to the Laplace operator on two-forms
by choosing  the gauge fixing condition
$$
\nabla^j h_{ji}-{1\over2} \nabla_i h^j{}_j=0
$$
 for infinitesimal diffeomorphisms and by 
using the relation $Y_{ij}=h_{ik} J^k{}_i$
 between symmetric (1,1)-tensors and (1,1)-forms. The Hermiticity
 condition for the metric implies that the deformations $h$ of the
 metric are (1,1) tensors.

Note that the above gauge can always be attained.
Suppose that $h$ does not satisfy the gauge.
Assume that there is an infinitesimal diffeomorphism $v$ such that
 $h'_{ij}=h_{ij}+\nabla_i v_j+\nabla_j v_i$
 satisfies the gauge. Then $v$ is determined by the equation
$$
\nabla^k \nabla_k v_i=\nabla^j h_{ji}-{1\over2} \nabla_i h^j{}_j~.
$$
To show that  there is always such a $v$, we must
invert  the above equation. This can be achieved iff
$\nabla^j h_{ji}-{1\over2} \nabla_i h^k{}_k$ is orthogonal to the kernel of the
elliptic operator $\nabla^k \nabla_k$. We rewrite the above equation as
$$
(d\delta+\delta d) V=-\nabla^j Y_{ji}+{1\over2} J^k{}_i \nabla_k h^j{}_j~,
$$
where $V_i= v_k J^k{}_i$ and $\delta$ is the adjoint of $d$.
Of course this equation can be inverted iff the right-hand-side is orthogonal
to harmonic one-forms.  Indeed let $Z$ be a harmonic one-form, then
$$
\int_M Z^i\left(-\nabla^j Y_{ji}+{1\over2} \Omega_{ki} \nabla^k h^j{}_j\right)
d{\rm vol}=
-\int_M \left(\nabla_i Z_j Y^{ij} +{1\over2} \nabla_i Z_j~ \Omega^{ij}~ h^k{}_k
\right) d{\rm vol}=0
$$
because $Z$ is closed.

 \appendix{Another derivation of the field equations for (2,0) compactifications}

 The first-order corrections $(h, f, \phi)$ to the background of a Calabi-Yau
 compactification $(g, \varphi)$, $T=0$, can be determined
 by the field equations without using
 the conditions for spacetime supersymmetry. To show
 this, we substitute (\ref{expeq}) in the
field equation (\ref{feq}) and collect the linear terms in $\alpha'$.
Using the fact that at zeroth order in $\alpha'$ the geometry is Calabi-Yau,
the equation for the metric gives
\beq
\Delta_Lh_{ij}+ S_{ij} +2\nabla_i \partial_j\phi=0~,
\label{eina}
\eeq
where $\Delta_L$ is the Lichnerowicz
operator (see appendix B).  In this case, we have
\beq
\Delta_Lh_{ij}=-{1\over2}\nabla^2 h_{ij}-
 R_{ik jl} h^{kl}+{1\over2}\nabla_i\nabla^kh_{kj}+
{1\over2}\nabla_j\nabla^k h_{ki}
-{1\over2}\nabla_i\nabla_j h^k{}_k
\label{lop}
\eeq
because $X_n$ is Calabi-Yau and $R_{ij}=0$,
and
\beq
S_{ij}={1\over4}[ R_{iklm} R_j{}^{klm}
- F_{ikab} F_j{}^{kab}]
\label{ann}
\eeq
The covariant derivative $\nabla$ in (\ref{eina}) and (\ref{lop})
and the curvature $R$ are  with respect to the Calabi-Yau metric $g$.

To solve equation (\ref{eina}) with respect to
$h$, we shall exploit the well-known relation between the
Lichnerowicz and Laplace operators
on Calabi-Yau manifolds. There are two ways to do this.
One is to use the scheme dependence
of the two-loop beta function and the other is to impose a gauge fixing condition
on the deformations $h$ of the metric $g$. The latter is common in moduli problems
in order for the deformation $h$ to be orthogonal to the orbits of infinitesimal
diffeomorphisms.
Using the scheme dependence of the two-loop beta function, we can 
arrange so that
the term
$$
{1\over2}\nabla_i\nabla^kh_{kj}+{1\over2}\nabla_j\nabla^kh_{ki}
+{1\over2} r \nabla_i \partial_j h^k{}_k
$$
is cancelled  by a wave function renormalization, where $r$ is a real number.
Alternatively, one can impose the gauge fixing condition
$$
\nabla^k h_{ki}+ {r\over 2} \nabla_i h^k{}_k=0~.
$$
Provided that we set for the deformation of the dilaton
$$
\phi={1\over 4} (r+1) h^k{}_k~,
$$
the remaining equation is
\beq
-{1\over2}\nabla^2 h_{ij}- R_{ik jl} h^{kl}+S_{ij}=0~.
\label{lap}
\eeq
To solve this equation observe that $S$ is a (1,1) symmetric tensor with respect to the
complex structure. This allows us to consider deformations of the metric which
are (1,1) as well, i.e. to take $h$ to be a (1,1) symmetric tensor.
It is well known that on K\"ahler manifolds there is a 1-1 correspondence between
symmetric (1,1) tensors and (1,1) two-forms. Indeed define the  forms
$Y_{ij}=h_{ik} J^k{}_j$ and $Z_{ij}=S_{ik} J^k{}_j$ associated to $T$ and $S$.
Equation (\ref{lap}) becomes
\beq
\Delta Y+Z=0
\label{lapa}
\eeq
where
$$
\Delta Y_{ij}=-{1\over2} \nabla^k \nabla_k Y_{ij}-R_{ikj\ell} Y^{k\ell}
$$
is the standard Laplace operator on $Y$ (see appendix B). We have
taken into account that Calabi-Yau manifolds are Ricci-flat, $R_{ij}=0$.
So it remains to invert the
Laplace operator to determine $Y$ in terms of $Z$.

To solve the equation (\ref{lapa}) in terms of $Y$, we have to show that
$Z$ is orthogonal to
the harmonic two-forms of $X_n$ in the Hodge decomposition with respect to the
Calabi-Yau metric $g$. This can be achieved by relating
the two-loop contribution $Z$ to the beta function of the metric, to the
heterotic anomaly $P$. Using
$$
\Omega^{ij} R_{ijkl}=\Omega^{ij} F_{ij ab}=0~,
$$
we can show that
$$
Z_{ij}={1\over4} P_{ijmn} \Omega^{mn}~.
$$
Now suppose that $n=3$. After some computation, we find that
$$
Z_{ij}=-{1\over 2} {}^*P_{ij}+{1\over 16} \Omega_{ij} P_{mnpq} \Omega^{mn} \Omega^{pq}~.
$$
The cancellation of the global anomaly implies that $P$ is exact.
As a consequence the dual two-form ${}^*P$ is co-exact and
so it is orthogonal to harmonic two-forms. It remains to show that
$P_{mnpq} \Omega^{mn} \Omega^{pq}$
is not harmonic.  Observe  that a harmonic function on $X_3$ is a non-vanishing constant.
Integrating the identity
$$
P\wedge \Omega={1\over8} P_{mnpq} \Omega^{mn} \Omega^{pq} d{\rm vol}
$$
over the compact manifold $X_3$ and using that $P$ is exact, it is easy
to see that $P_{mnpq} \Omega^{mn} \Omega^{pq}$ is not harmonic.

This result can be easily extended to four- and eight-dimensional
internal manifolds.  The $n=2$ case has been
investigated in \cite{hpap}. For
$n=4$, the computation is similar to $n=3$. The only difference
is the relation between the two-loop counterterm and the anomaly.

Next we  investigate the field equation of the three-form field strength.
Again substituting (\ref{expeq}) in (\ref{feq}) and
collecting the terms linear in $\alpha'$, we find
$$
\nabla^i f_{ijk}=0\
$$
or equivalently
$$
d^\dagger f=0~,
$$
where $d^\dagger$ is the adjoint operator of $d$.
To derive this, we have used the fact that $H$ vanishes at
zeroth order  in $\alpha'$. In addition, we have that  $df=-P$; $f$
is well-defined because $P$ is exact.
Using the Hodge decomposition, we can write
$$
f=f_h+dX+d^\dagger W
$$
where $f_h$ is harmonic and $W$ is a four-form.
Adding the counterterm
$$
b=-\alpha' X\ ,
$$
the exact three-form $dX$ can be eliminated as follows
$$
H=-\alpha' dX+\alpha' f= \alpha' f_h+\alpha' d^\dagger W~.
$$
Thus the field equation is satisfied and $dH=-P$.
Alternatively, one can choose $f$ such that $f=f_h+d^\dagger W$.



\end{document}